\documentclass[12pt,a4paper]{article}
\usepackage{amsfonts,latexsym}
\usepackage{amsmath,amssymb}
\usepackage{graphicx,color}

\renewcommand{\thefootnote}{\fnsymbol{footnote}}

\begin{document}

\vspace{12mm}
\begin{center}
{{{\Large {\bf Charged  qOS-extremal black hole and its scalarization by entropy function approach }}}}\\[10mm]
{Yun Soo Myung\footnote{e-mail address: ysmyung@inje.ac.kr}}\\[8mm]

{Center for Quantum Spacetime, Sogang University, Seoul 04107, Republic of  Korea\\[0pt] }

\vspace{12mm}

\end{center}
 \begin{abstract}
    \noindent We  investigate  scalarization of charged quantum Oppenheimer-Snyder extremal (cqOSe)-black hole in the Einstein-Gauss-Bonnet-scalar theory with a nonlinear electrodynamics term.
    This black hole is described by  quantum parameter $\alpha$  and magnetic  charge $P$.  It is equivalent to the qOS-extremal black hole whose action is still unknown  when  imposing a relation of $(3\alpha P^2)^{1/4}\to 3M/2$.   Focusing on the onset of scalarization, we find the single branch of scalarized cqOS extremal (scqOSe)-black holes. To obtain a scalar cloud (seed) for the single branch, however, we have to consider its  near-horizon geometry of the Bertotti-Bobinson (BR) spacetime. In this case, two scalar clouds for positive and negative coupling constant $\lambda$ are found to represent two branches.
    Applying Sen's entropy function approach to this theory, we obtain the entropy which is  the only physical quantity to describe the scqOSe-black holes.
    We find that the positive branch is preferred than  the negative branch.
\end{abstract}

\vspace{1.5cm}

\hspace{11.5cm}{Typeset Using \LaTeX}
\newpage
\renewcommand{\thefootnote}{\arabic{footnote}}
\setcounter{footnote}{0}

\vspace{2mm}

\section{Introduction}

The quantum Oppenheimer-Snyder (qOS)-black hole was recently found by considering the qOS model in  loop quantum cosmology~\cite{Lewandowski:2022zce}.
The qOS model describes a collapsing  dust ball inside the qOS-black hole, while the quantum Swiss Cheese model describes the qOS-black hole  surrounded by Ashtekar-Pawlowski-Singh model~\cite{Ashtekar:2006rx}.
However, one does not know  its action $\mathcal{L}_{\rm qOS}$ to give the qOS-black hole described by mass ($M$) and quantum parameter ($\alpha$) as a direct solution.
Regarding to  appropriate actions for $\mathcal{L}_{\rm qOS}$,  a candidate was proposed  by introducing   a nonlinear electrodynamics (NED) term  $\mathcal{L}_{\rm NED}$~\cite{Mazharimousavi:2025lld}.
In this case, one has found charged quantum Oppenheimer-Snyder(cqOS)-black hole and  the qOS-black hole can be  recovered when choosing  $P=M$.
Various aspects of the qOS-black hole have been  studied by including   quasinormal mode analysis for tensor and scalar perturbations~\cite{Skvortsova:2024msa}, thermodynamics~\cite{Dong:2024hod,Panotopoulos:2025ygq}, shadow radius~\cite{Yang:2022btw,Ye:2023qks}, and scalarization within the Einstein-Gauss-Bonnet-scalar (EGBS) theory~\cite{Chen:2025wze,Myung:2025pmx}.

On the other hand, extremal black holes have  played a crucial role in various aspects. They possessing
zero Hawking temperature and heat capacity, are expected to bring us valuable insights into  black hole
thermodynamics~\cite{Charmousis:2010zz} and  Hawking radiation~\cite{Angheben:2005rm}. For these black holes, hence,  the entropy is regarded as the only physical quantity.
In the astrophysical aspect,  many astrophysical black holes are supposed  to be  nearly extremal~\cite{Volonteri:2004cf,Gou:2013dna}.
To understand  the nature of extremal black holes, one has   to study the dynamical properties
of test fields  propagating around them but it  requires  the investigation to confine  to  the near-extremal limit
or the near-horizon approximation.

In addition, the no-hair theorem states  that a black hole can be completely described  by  three  parameters: mass ($M$), electric charge ($Q$), and rotation parameter ($a$) in Einstein-Maxwell gravity
\cite{Carter:1971zc,Ruffini:1971bza}. If a scalar field is minimally coupled to  gravitational and electromagnetic  fields, it turned out that there is no black hole solution with scalar hair~\cite{Herdeiro:2015waa}.
However, its evasion  occurred frequently in the context of scalar-tensor theories with  the nonminimal scalar coupling  either to Gauss-Bonnet (GB) term~\cite{Doneva:2017bvd,Silva:2017uqg,Antoniou:2017acq} or to Maxwell term~\cite{Herdeiro:2018wub,Myung:2018vug}.  The former is called GB$^+$ scalarization  with a positive coupling parameter and the latter is called M$^+$ scalarization, both were triggered by tachyonic instability.
Here, one found infinite branches of scalarized black holes induced by infinite scalar clouds.
For review on spontaneous scalarization, one may see Ref.~\cite{Doneva:2022ewd}.

The GB$^-$ scalarization can be  realized for the black hole with double horizons when coupling $f(\phi)$ to the GB term with negative coupling parameter.
It was shown that  two branches of positive ($\gamma>0$) and negative ($\gamma<0$) coupling parameter  are allowed  for  scalarization of the Reissner-Nordstr\"om (RN) black holes  in the Einstein-Gauss-Bonnet-Naxwell-scalar (EGBMS) theory with coupling function $f(\phi)=\gamma \phi^2$~\cite{Brihaye:2019kvj}.  It was  claimed that  the appearance of  negative branch is related to  its near-horizon geometry (AdS$_2\times S^2$) of extremal RN black hole.
However, the negative  branch appeared from the sign change of GB term~\cite{Herdeiro:2021vjo}. For qOS-extremal black holes, its scalar potential has a negative region in the whole near-horizon~\cite{Myung:2025oik}. This is the origin of GB$^-$ scalarization for charged black holes with negative coupling constant, inducing the single branch of scalarized black holes.
At this stage, one has to realize that it is not easy to obtain its scalar cloud,  which may be  a seed to generate the single branch of scalarized extremal black holes.
This is because numerical methods cannot solve the linearized scalar equation to find out scalar clouds in the extremal black hole background~\cite{Richartz:2015saa,Senjaya:2025pyv}.
It  forces the numerical investigation to end at the near-extremal limit~\cite{Hod:2017gvn} or at the near-horizon approximation.
To obtain the  scalar clouds, here,  one  may use   the near-horizon geometry [Bertotti-Bobinson (BR: AdS$_2\times S^2$) spacetime] of extremal black holes.

In the present work, we  wish to   investigate GB$^-$ scalarization of cqOSe-black hole in the EGBS theory with the NED term $\mathcal{L}_{\rm NED}$ and  the coupling function $f(\phi)=2(\lambda \phi^2-\zeta \phi^4)$ to GB term.
Here, $\lambda$ and $\zeta$ represent two scalar coupling parameters.
    This extremal black hole is described by two of  quantum parameter $\alpha$  and magnetic charge $P$.  It is equivalent to the qOS-extremal black hole whose action ($\mathcal{L}_{\rm qOS}$) is still unknown  when  imposing a relation of $(3\alpha P^2)^{1/4}\to 3M/2$. This model suggests an alternative to studying scalarization of qOS-extremal black hole.  Studying on the onset of scalarization based on the linearized theory with $\lambda$, we find the single branch of scalarized cqOS extremal (scqOSe)-black holes. To obtain a scalar cloud (seed), we consider its  near-horizon geometry of the BR spacetime. Two scalar clouds are found to represent two branches.
    Applying Sen's entropy function approach~\cite{Sen:2005wa} to this theory, we obtain the entropy function which indicates  the only physical quantity to describe the scqOSe-black holes.

\section{cqOS-extremal black hole}
The EGBS theory with the nonlinear electrodynamics (NED) term takes the form~\cite{Chen:2025wze,Myung:2025pmx}    as
\begin{equation}
\mathcal{L}_{\rm EGBS-N}=\frac{1}{16\pi}\Big[ R-2\partial_\mu \phi \partial^\mu \phi+ f(\phi) {\cal R}^2_{\rm GB}+{\cal L}_{\rm NED}\Big],\label{Action1}
\end{equation}
where $\phi$ is the scalar field and   $f(\phi)=2(\lambda \phi^2-\zeta \phi^4)$ denotes a quartic coupling function with   $\lambda$ and $\zeta$  scalar coupling parameters.
Also, ${\cal R}^2_{\rm GB}=R^2-4R_{\mu\nu}R^{\mu\nu}+R_{\mu\nu\rho\sigma}R^{\mu\nu\rho\sigma}$ denotes the GB term  while  ${\cal L}_{\rm NED}=-2\xi(\mathcal{F})^{3/2}$ represents the NED term with $\xi=\frac{3\alpha}{2^{3/2} P}$ and $\mathcal{F}=F_{\mu\nu}F^{\mu\nu}$ the Maxwell term.

The Einstein  equation  with  $G_{\mu\nu}=R_{\mu\nu}-(R/2)g_{\mu\nu}$ is derived as
\begin{eqnarray}
 G_{\mu\nu}=2\partial _\mu \phi\partial _\nu \phi -(\partial \phi)^2g_{\mu\nu}+\Gamma_{\mu\nu}^\phi+T^{\rm NED}_{\mu\nu}, \label{equa1}
\end{eqnarray}
where $\Gamma_{\mu\nu}^\phi$ with $\Psi_{\mu}= f'(\phi)\partial_\mu \phi$  is given by
\begin{eqnarray}
\Gamma_{\mu\nu}^\phi&=&2R\nabla_{(\mu} \Psi_{\nu)}+4\nabla^\alpha \Psi_\alpha G_{\mu\nu}- 8R_{(\mu|\alpha|}\nabla^\alpha \Psi_{\nu)} \nonumber \\
&+&4 R^{\alpha\beta}\nabla_\alpha\Psi_\beta g_{\mu\nu}
-4R^{\beta}_{~\mu\alpha\nu}\nabla^\alpha\Psi_\beta  \label{equa2}
\end{eqnarray}
and the NED energy-momentum tensor takes the form
\begin{equation}
T^{\rm NED}_{\mu\nu}=6\xi \Big[\sqrt{\mathcal{F}}F_{\mu\lambda}F_\nu^\lambda-\frac{1}{6}\mathcal{F}^{3/2}g_{\mu\nu}\Big]. \label{EM-t}
\end{equation}

Considering  $\mathcal{F}=2P^2/r^4$ for a magnetically charged configuration ($F_{\theta\varphi}=P\sin \theta$), its energy-momentum tensor takes the form
\begin{equation}\label{q-em}
T^{\rm NED, \nu}_{\mu}=\frac{3\alpha P^2}{r^6}{\rm diag}[-1,-1,2,2].
\end{equation}
In this case, choosing $P=M$ leads to the energy-momentum tensor for qOS-black hole~\cite{Lewandowski:2022zce}
\begin{equation}
T^{\rm qQS, \nu}_{\mu}=\frac{3\alpha M^2}{r^6}{\rm diag}[-1,-1,2,2],
\end{equation}
whose matter action is still unknown because this  was derived by making use of the junction conditions.
The scalar field equation is given by
\begin{equation}
\square \phi +\frac{1}{4}f'(\phi) {\cal R}^2_{\rm GB}=0 \label{s-equa}.
\end{equation}
Taking into account $G_{\mu\nu}=T^{\rm NED}_{\mu\nu}$ together  with $\phi=0$,  the cqOS-black hole  solution is obtained as~\cite{Mazharimousavi:2025lld}
\begin{equation} \label{ansatz}
ds^2_{\rm cqOS}= \bar{g}_{\mu\nu}dx^\mu dx^\nu=-g(r)dt^2+\frac{dr^2}{g(r)}+r^2d\Omega^2_2
\end{equation}
whose metric function is defined  by the mass function $m(r)$ as
\begin{equation}
g(r,M,\alpha,P)\equiv 1-\frac{2m(r)}{r}=1-\frac{2M}{r}+\frac{\alpha P^2}{r^4}. \label{g-sol}
\end{equation}
Here, the quantum parameter is given by $\alpha=16\sqrt{3}\pi\gamma^3$ with $\gamma$ the dimensionless Barbero-Immirzi parameter. For $\gamma=0.2375$, one finds that $\alpha=1.1663$~\cite{Meissner:2004ju,Domagala:2004jt}.

One finds two real solutions and complex solutions from $g(r)=0$
\begin{eqnarray}
&&r_i(M,\alpha,P),~{\rm for}~i=1,2,3,4, \label{f-roots}
\end{eqnarray}
where $r_{1}$ and $r_{2}$ become complex solutions, whereas $r_3(M,\alpha,P)\to r_-(M,\alpha,P)$ and $r_4(M,\alpha)\to r_+(M,\alpha,P)$.
Their  explicit forms  are given by
\begin{eqnarray}
r_\pm(M,\alpha,P)&=&\frac{M}{2}+\frac{1}{2}\Big[M^2+\frac{2^{5/3}P^2\alpha}{(3\eta)^{1/3}}+ \frac{(2\eta)^{1/3}}{3^{2/3}}\Big]^{1/2} \nonumber \\
&\pm& \frac{1}{2}\Big[2M^2-\frac{2^{5/3}P^2\alpha}{(3\eta)^{1/3}}- \frac{(2\eta)^{1/3}}{3^{2/3}}+\frac{2M^3}{(M^2+\frac{2^{5/3}P^2\alpha}{(3\eta)^{1/3}}+ \frac{(2\eta)^{1/3}}{3^{2/3}})^{1/2}}\Big]^{1/2}
 \label{oi-hor}
\end{eqnarray}
with
\begin{equation}
\eta(M,\alpha,P)=\alpha P^2\Big(9M^2+\sqrt{3}\sqrt{27M^4-16\alpha P^2}\Big). \label{eta-eq}
\end{equation}
From Eq.(\ref{eta-eq}), one obtains  a condition for the existence of  two horizons as
\begin{equation}
0<\alpha<\frac{27M^4}{16P^2},
\end{equation}
which leads to a cqOS-extremal black hole for $\alpha=\frac{27M^4}{16P^2}$ as
\begin{equation}
g_e(r,\alpha,P)=1-\frac{4(\alpha P^2)^{1/4}}{3^{3/4} r}+\frac{\alpha P^2}{r^4} \to \Big(1-\frac{(3\alpha P^2)^{1/4}}{r}\Big)^2\Big(1+\frac{2(\alpha P^2)^{1/4}}{3^{3/4}r}+\frac{\sqrt{\alpha P^2}}{\sqrt{3}r^2}\Big).
\end{equation}
Considering the relation  $\frac{2(3\alpha P^2)^{1/4}}{3}=M$, one finds the qOS-extremal black hole solution as
\begin{equation}
g_{\rm qOSe}(r,M)=1-\frac{2M}{r}+\frac{27M^4}{16 r^4}\to \Big(1-\frac{3M}{2r}\Big)^2\Big(1+\frac{M}{r}+\frac{3M^2}{4r^2}\Big).
\end{equation}
Here, we present  two extremal  horizons  from $g_e(r)=0$ and $g_{\rm qOSe}(r)=0$ as
\begin{equation}
 r_e(\alpha, P)=(3\alpha P^2)^{1/4},\quad r_e(M)=\frac{3M}{2}.
\end{equation}
This shows that two extremal representations are equivalent to each other.
Also, from $m(r)=0$, one finds the bounce radius
\begin{equation}
r_b=\Big(\frac{\alpha P^2}{2M}\Big)^{1/4},
\end{equation}
which is located inside the inner horizon $r_-$ (see Fig. 1(a)).

\begin{figure}
\centering
 \mbox{
   (a)
  \includegraphics[width=0.4\textwidth]{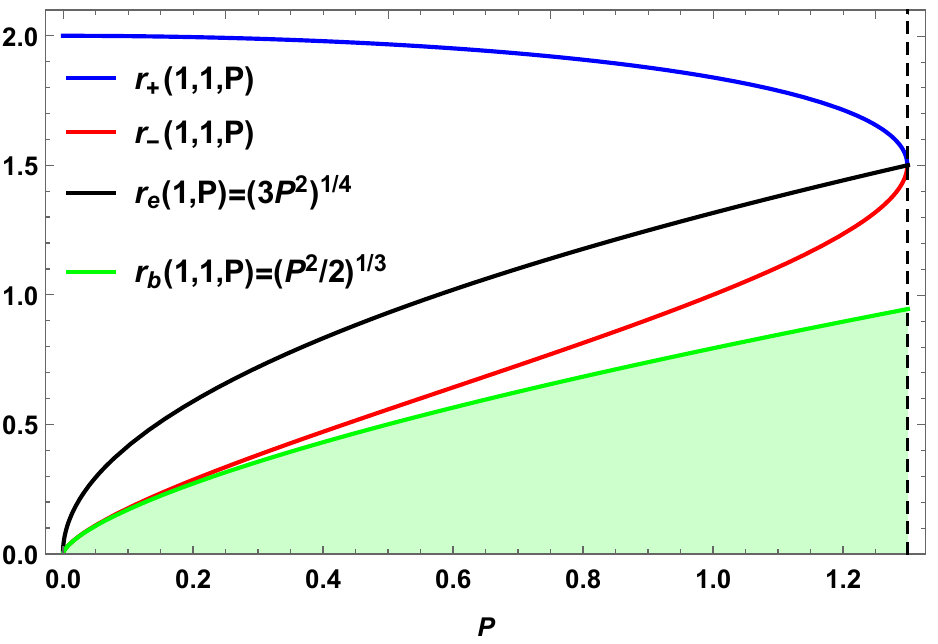}
 (b)
    \includegraphics[width=0.4\textwidth]{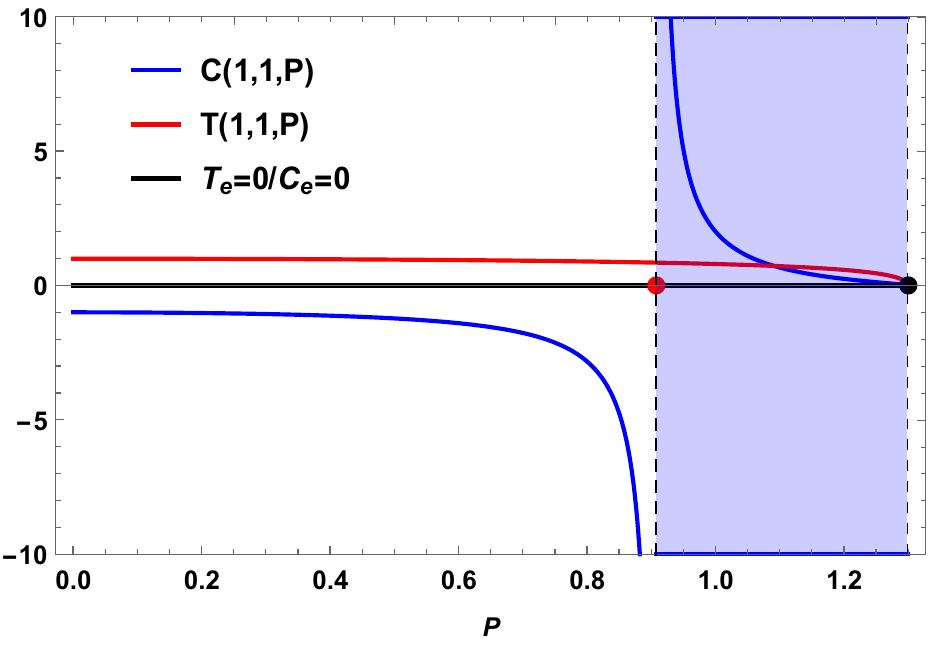}}
\caption{ (a) Two  horizons  $r_{\pm}(M=1,\alpha=1,P)$ are   function of   $P\in[0,P_e=1.299]$, showing the upper bound for the charge $P$.  Here, $r_e(\alpha=1,P)$ as a function of $P$ represents the extremal horizon, starting from $P=0$. The bounce radius $r_b(M=1,\alpha=1,P)$ is located inside the inner horizon. (b) Heat capacity $C(M=1,\alpha=1,P)/|C_S(1,0)|$ with $|C_S(1,0,0)|=25.13$.  Heat capacity blows up at Davies point ($P_D=0.9077,$ red dot) where the temperature $T(1,1,P)$ has the maximum.   The shaded region denotes $C>0$ for cqOS-black holes. The heat capacity and temperature are always zero at the cqOSe-black line.}
\end{figure}

As is shown in Fig. 1(a), there exist outer/inner horizons  $r_\pm(M=1,\alpha=1,P)$ as functions of $P$  with  the upper bound at $P=P_e$ for the magnetic charge  of cqOSe-black hole.
Also, we display the extremal horizon $r_e(\alpha=1,P)$ as a function of $P$ and the bounce radius $r_b(M=1,\alpha=1,P)$ is located inside the inner horizon.

The temperature $T=\frac{\partial m}{\partial S}$ and  heat capacity $C=\frac{\partial m}{\partial r_+}(\frac{\partial T}{\partial r_+})^{-1}$ with mass $M\to m(M,\alpha,P)=(\alpha P^2 +r_+^4)/2r_+^3$ and area-law entropy $S=\pi r_+^2$ are given by
\begin{eqnarray}
T(M,\alpha,P)&=& \frac{-3\alpha P^2 +r_+^4(M,\alpha,P)}{ 4 \pi r_+^5(M,\alpha,P)}, \label{tem1} \\
 C(M,\alpha,P)&\equiv& \frac{NC(M,\alpha,P)}{DC(M,\alpha,P)}=-\frac{2\pi r_+^2(M,\alpha,P)[-3\alpha P^2 +r_+^4(M,\alpha,P)]}{-15\alpha P^2+r_+^4(M,\alpha,P)}.                         \label{heat1} \\
\end{eqnarray}
Here, the Davies point (blow-up point) can be obtained from solving $DC(M,\alpha,P)=0$.
We observe from Fig. 1(b) that  $C(1,1,P)/|C_S(1,0,0)|$ blows up at Davies point ($P_D=0.9077$, red dot) where the temperature $T(M=1,\alpha=1,P)$ takes  the maximum value.
 The cqOS black hole is thermodynamically stable if $C>0 ~(P_D<P<P_e)$, while it is unstable for $C<0~(0<P<P_D)$. Hence, the Davies point is regarded as a critical point which  can represent a sharp phase transition from $C>0$ to $C<0$. However, it is important to note that  the temperature and heat capacity of cqOSe-black hole  are always zero as is shown in the black line (Fig.1(b)), implying that the entropy survives  the only physical quantity.
 \begin{figure}
\mbox{
   (a)
  \includegraphics[width=0.38\textwidth]{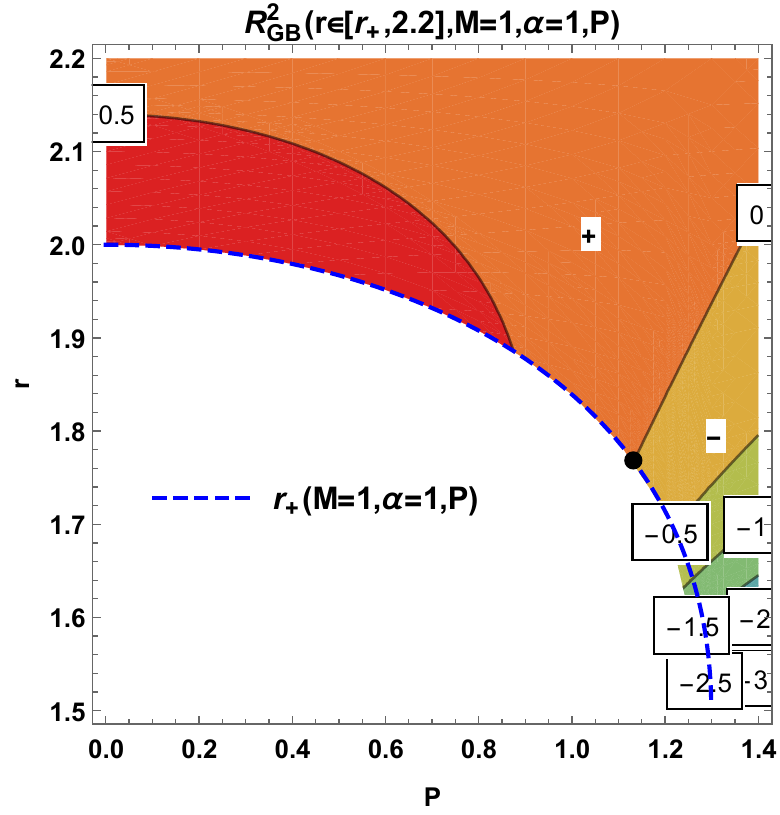}
 (b)
    \includegraphics[width=0.38\textwidth]{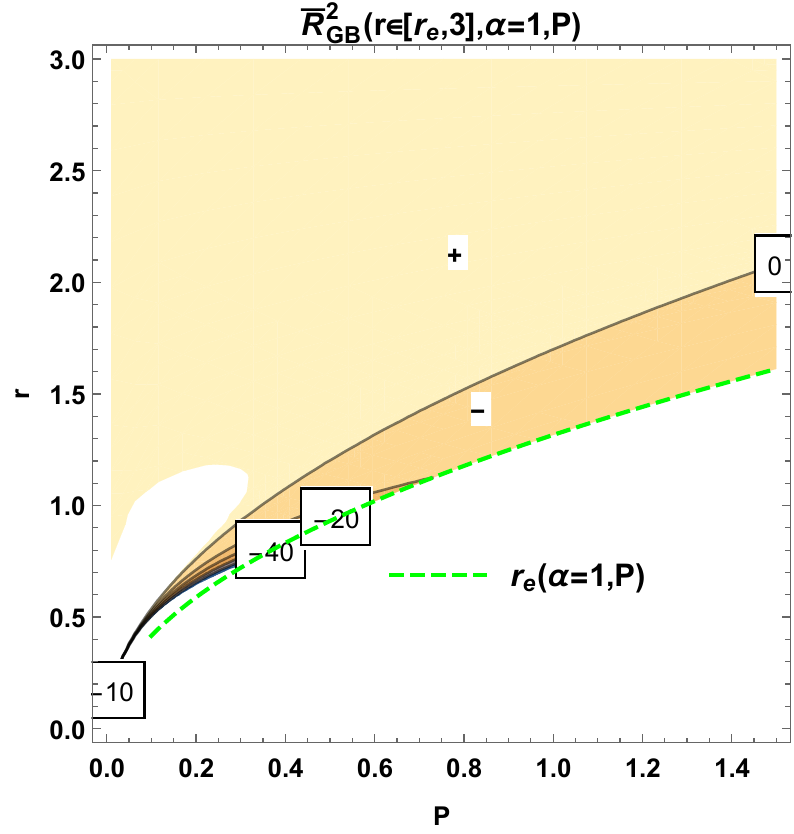}}
\caption{ (a) $\mathcal{R}^2_{\rm GB}(r,M=1,\alpha=1,P)<0$  as  a function of
$(r\in [r_+(M=1,\alpha=1,P),2.2],P\in[0,1.5])$ in the near-horizon for cqOS black holes. Its sign changes from positive ($0<P<P_c~(=1.1329)$) to negative ($P_c<P<P_e~(=1.299)$) with the critical onset point $[(P_c,r_+(1,1,P_c)$, black dot].
 (b)  $\bar{\mathcal{R}}^2_{\rm GB}(r,\alpha=1,P)<0$  as  a function of
$(r\in [r_e(\alpha=1,P),3],P\in[0,1.5])$ in the whole near-horizon for cqOSe-black holes. Hence, one finds that $-\lambda \bar{\mathcal{R}}^2_{\rm GB} <0$ in the whole near-horizon for $\lambda <0$, leading to tachyonic instability.  }
\end{figure}

For cqOS black hole background, the GB term is given by
\begin{equation}
\mathcal{R}^2_{\rm GB}=\frac{48 M^2}{r^{6}}\Big[\frac{3\alpha^2P^4}{M^2 r^6}-\frac{5\alpha P^2}{M r^3} +1\Big].
\end{equation}
Th near-horizon behavior for $\mathcal{R}^2_{\rm GB}$ is depicted in Fig. 2(a).
Its sign changes from positive [$0<P<P_c~(=1.1329)$] to negative [$P_c<P<P_e~(=1.299)$] with the critical onset point $[(P_c,r_+(1,1,P_c))$, black dot].
This implies that GB$^-$ scalarization is allowed only  for $\lambda<0$. It indicates that  tachyonic instability of $-\lambda \bar{\mathcal{R}}^2_{\rm GB}<0$ is available for $P_c<P<P_e$ (narrow strip) in the near-horizon.
This corresponds to the origin for  GB$^-$ scalarization of cqOS black holes with negative $\lambda$.

\section{Onset scalarization}
This analysis is based on the linearized scalar theory described by
\begin{equation}
(\bar{\square}+\lambda \bar{\mathcal{R}}^2_{\rm GB}) \delta \phi=0.
\end{equation}

\subsection{GB$^{\rm e}$ scalarization for cqOSe-black hole}

In this section, we wish to focus on the GB$^{\rm e}$ onset scalarization of the cqOSe-black holes.
Here, we use $g_e(r,\alpha,P)$ instead of $g(r,M,\alpha,P)$, implying that it includes the degenerate  horizon.
In this case, its spacetime is described by
\begin{equation}
ds^2_{\rm e}=-g_e(r)dt^2+\frac{dr^2}{g_e(r)}+r^2d\Omega^2_2,
\end{equation}
which possesses an AdS$_2 \times S^2$ as the near-horizon geometry.

Let us introduce the tortoise coordinate $r_*$  defined as $ dr_*=dr/g_e(r)$ and consider the separation of variables
\begin{equation}
\delta \phi(t,r_*,\theta,\varphi)=\sum_{m} \sum_{l=|m|}^\infty \frac{\psi_{lm}(t,r_*)}{r}Y_{lm}(\theta,\varphi).
\end{equation}
Here, the radial part of the ($l=0,m=0)$-linearized scalar equation  takes the form
\begin{equation} \label{emode-d}
\frac{\partial^2\psi_{00}(t,r_*)}{\partial r_*^2} -\frac{\partial^2\psi_{00}(t,r_*)}{\partial t^2}=V_e(r)\psi_{00}(t,r_*),
\end{equation}
where  the scalar potential $V_e(r)$ is given by
\begin{equation} \label{epot-c}
V_e(r,\alpha,P,\lambda)=g_e(r,\alpha,P)\Big[\frac{4(\alpha P^2)^{1/4}}{3^{3/4}r^3}-\frac{4\alpha P^2}{r^6}+\tilde{m}^2_e\Big]
\end{equation}
with its effective mass term
\begin{equation}
\tilde{m}^2_e\equiv -\lambda \bar{\mathcal{R}}^2_{\rm GB}=-\frac{16\lambda }{3r^{12}}\Big[27 \alpha^2P^4-30 \cdot  3^{1/4}(\alpha P^2)^{5/4}r^3+4\sqrt{3\alpha P^2} r^6\Big]. \label{e-masst}
\end{equation}
As is shown in Fig. 2(b), the GB term  $\bar{\mathcal{R}}^2_{\rm GB}(r,\alpha=1,P)$ is regarded  as a function of
($r\in [r_+(\alpha=1,P),3],P\in[0,1.5])$. Here, one observes  that $\tilde{m}^2_e <0$ in the whole  near-horizon for $\lambda <0$, leading to tachyonic instability.
\begin{figure}
  \mbox{
   (a)
  \includegraphics[width=0.4\textwidth]{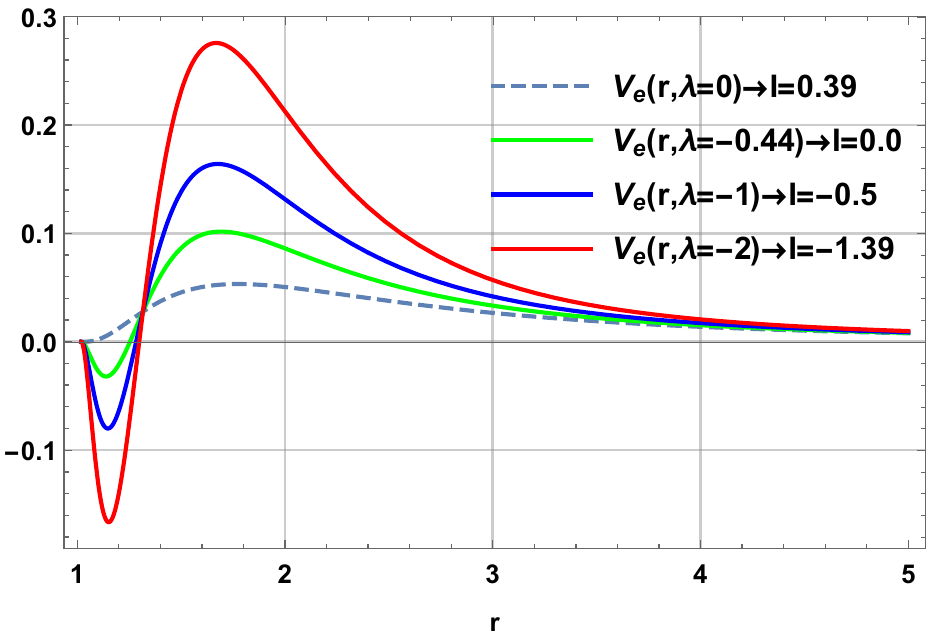}
 (b)
    \includegraphics[width=0.4\textwidth]{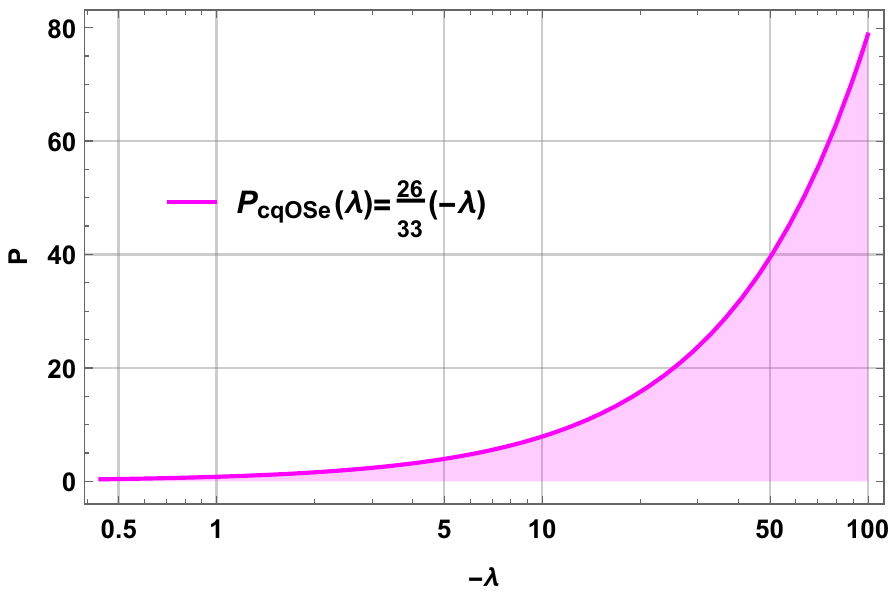}}
\caption{ (a) Extremal scalar potentials $V_e(r,\alpha=1,P=0.6,\lambda)$ with $\lambda=0,-0.44,-1,-2$ as a function of $r\in[r_+=1.019,5]$ for GB$^{\rm e}$ scalarization. Here, one has the integration $I=0.39(\lambda=0),~0(\lambda=-0.44),~-0.5(\lambda=-1),~-1.39(\lambda=-2)$. (b) Sufficient unstable (shaded) region of  cqOSe-black hole for GB$^{\rm e}$ scalarization. A single branch for $\lambda<0$ is allowed for $0<P<P_{\rm cqOSe}~(=-0.7879\lambda)$.  }
\end{figure}
Actually, Fig. 3(a) shows that the extremal potentials have negative regions  in the whole  near-horizon for any negative $\lambda$.

To obtain  a sufficient condition for the instability, one may use the  condition for instability proposed  by Ref.~\cite{Dotti:2004sh}
\begin{equation}
\int^\infty_{r_e(\alpha,P)}\Big[\frac{V_e(r,\alpha,P,\lambda)}{g_e(r,\alpha,P)}\Big] dr \equiv I(\alpha,P,\lambda)  <0. \label{sc-ta}
\end{equation}
This condition  leads to
\begin{equation}
I(\alpha,P,\lambda)=\frac{2(11\cdot 3^{3/4}\sqrt{\alpha P^2}+26 \cdot 3^{1/3} \lambda)}{165 (\alpha P^2)^{3/4}}<0,
\end{equation}
which is  solved for $P>0$ and $\lambda<0$ as
\begin{equation}
0<P\le P_{\rm cqOSe}(\lambda),\quad P_{\rm cqOSe}(\lambda)=-\frac{26\lambda}{33 \alpha}.
\end{equation}
The sufficiently unstable region is depicted by  the shaded region in Fig. 3(b).
However, one could not obtain scalar cloud which is a seed to  generate scalarized cqOSe-black holes.
This is because its boundary potential~\cite{Hod:2017gvn}
\begin{equation}
V_{\rm eb}(r,\alpha,P)=\frac{4}{\sqrt{3}r^6}  \sqrt{\frac{-27 (\alpha P^2)^2+30 \cdot  3^{1/4}(\alpha P^2)^{5/4}r^3-4\sqrt{3\alpha P^2} r^6}{g_e(r,\alpha,P)}},
\end{equation}
 is not properly defined for outside the outer horizon [$r>1.316>r_+(=1.019)]$ for $\lambda<0$.
This requires our  investigation to confine  to  the near-extremal limit
   or the near-horizon approximation.

\subsection{GB$^{\rm BR}$ scalarization to find scalar clouds}

In the previous section, we did not obtain  a numerical scalar cloud which might be  a scalar seed for scalarized cqOSe-black hole in the single branch.
Here, we wish to find   analytic  scalar clouds.
For the cqOSe-black hole, one always finds its near-horizon geometry of the BR background (AdS$_2\times S^2$) as~\cite{deCesare:2024csp}
\begin{equation}
ds_{\rm BR}^2=\frac{2}{g''_e(r_e)}\Big(-\rho^2d\tau^2+\frac{d\rho^2}{\rho^2}\Big) +r^2_e(d\theta^2+\sin^2\theta d\varphi^2), \label{ads-S}
\end{equation}
where
\begin{equation}
\frac{2}{g''_e(r_e)}=\frac{(3\alpha P^2)^{1/4}}{2}, \quad r_e^2=(3\alpha P^2)^{1/4}.
\end{equation}
Here, two coordinates ($\tau,\rho$) are dimensionless and  the extremal horizon is located at $\rho=0$.

Using  Eq.(\ref{ads-S}), we  compute  the GB term to define its mass term ($\mu^2$) as
\begin{equation}
-\lambda\mathcal{R}^2_{\rm GB}= \frac{16\lambda}{3\alpha P^2}\to \mu^2
\end{equation}
Then,  $s$-mode linearized equation for $\delta\phi(\tau,\rho)$
is given by
\begin{equation}
-\frac{1}{\rho^2}\partial^2_\tau\delta \phi+\partial_\rho(\rho^2\partial_\rho \delta \phi)-\mu^2 \delta \phi=0. \label{KG-eq}
\end{equation}
Introducing a tortoise coordinate $\rho_*=1/\rho$,
the $s$-mode scalar equation leads to~\cite{Brihaye:2019kvj}
\begin{equation} \label{ads-KGe}
 -\frac{\partial^2\delta \phi(\tau,\rho_*)}{\partial \tau^2}+\frac{\partial^2\delta \phi(\tau,\rho_*)}{\partial \rho_*^2}=V_{\rm GB}(\rho_*,\lambda) \delta \phi(\tau,\rho_*),
\end{equation}
where the GB potential is given by
\begin{equation}
V_{\rm GB}(\rho_*,\lambda)=\frac{\mu^2}{\rho_*^2}\to V_{\rm GB}(\rho,\lambda)=\mu^2 \rho^2.
\end{equation}
Now, we would like to mention the Breitenlohner-Freedman (BF) bound for a massive scalar propagating around AdS$_2$ spacetime~\cite{Breitenlohner:1982jf,Breitenlohner:1982bm}
\begin{equation}
\mu^2\ge \mu^2_{\rm BF}=-\frac{1}{4}.
\end{equation}
For $\mu^2<\mu_{\rm BF}^2$ case, it corresponds to tachyon propagating around  AdS$_2$ spacetime and this spacetime becomes unstable.

To obtain scalar clouds, it is important to solve the static scalar equation with $\omega=0$
\begin{equation}
\frac{\partial^2\delta \phi(\rho_*)}{\partial \rho_*^2} -V_{\rm GB}(\rho_*,\lambda)\delta \phi(\rho_*)=0.
\end{equation}
One finds a scalar cloud
\begin{equation}
\delta \phi(\rho_*,\lambda)=c_1(\rho_*)^{\frac{1}{2}+\nu}+c_2(\rho_*)^{\frac{1}{2}-\nu} \to\delta \phi(\rho,\lambda)=c_1(\rho)^{-\nu-\frac{1}{2}}+c_2(\rho)^{\nu-\frac{1}{2}} \label{scl-eq}
\end{equation}
with
\begin{equation}
\nu=\frac{\sqrt{4\mu^2+1}}{2}.
\end{equation}

Choosing $\mu^2=-8(\lambda=-3\alpha P^2/2=-1.01)$ and $c_1=c_2=1/2$, the tachyonic  seed and its potential are given by
\begin{equation}
\delta \phi(\rho,\mu^2=-8)=\frac{1}{\sqrt{\rho}}\cos\Big[\frac{\sqrt{31} \ln(\rho)}{2}\Big], \quad V_{\rm GB}(\rho,\mu^2=-8)=-8  \rho^2, \label{tach-s}
\end{equation}
which  has many nodes as is shown Fig. 4(a), but it takes a large value of 100 at $\rho=10^{-4}$.  This tachyonic cloud represents one (negative) branch  for  scalarization of cqOSe-black holes.
To compare it with the conventional scalar clouds, one introduce  regular (finite at the horizon)  scalar clouds labelled by  number of nodes ($n=0,1,2,\cdots$) for GB$^+$ scalarization of  Schwarzschild black holes
[see Fig. 4(b)]. These  were obtained  by solving static linearized equation numerically~\cite{Myung:2018iyq}:
 $\delta \phi_0(r,M=1)$ has zero node (zero crossing at $r$-axis) with $n=0$ branch point $\lambda_0=0.73$, $\delta \phi_1(r,M=1)$ has one node with $n=1$ branch point $\lambda_1=4.87$,  and $\delta \phi_2(r,M=1)$ has two nodes with $\lambda_2=12.8$.

Selecting  $\mu^2=8(\lambda=3\alpha P^2/2=1.01)$ with $c_1=1$ and  $c_2=0$, one finds from  Eq.(\ref{scl-eq}) as
 \begin{equation}
 \delta \phi(\rho,\mu^2=8)=\rho^{-\frac{1}{2}(\sqrt{33}+1)}, \label{inf-sc}
 \end{equation}
 which shows that $\delta \phi(\rho,\mu^2=8)$  approaches infinity as $\rho\to 0$, but it is zero at $\rho=\infty$ without node (see Fig. 4(c)) and thus, it is called the blow-up scalar cloud at the horizon even though it is similar to the $n=0$ branch  in Fig. 4(b). This can represent the other (positive) branch.
In addition,   we note that the other form of $\rho^{\frac{1}{2}(\sqrt{33}-1)}$ approaches zero as $\rho\to 0$ while it takes the infinity as $\rho \to \infty$, corresponding to a non-normalizable solution.
  \begin{figure}
\mbox{
   (a)
  \includegraphics[width=0.27\textwidth]{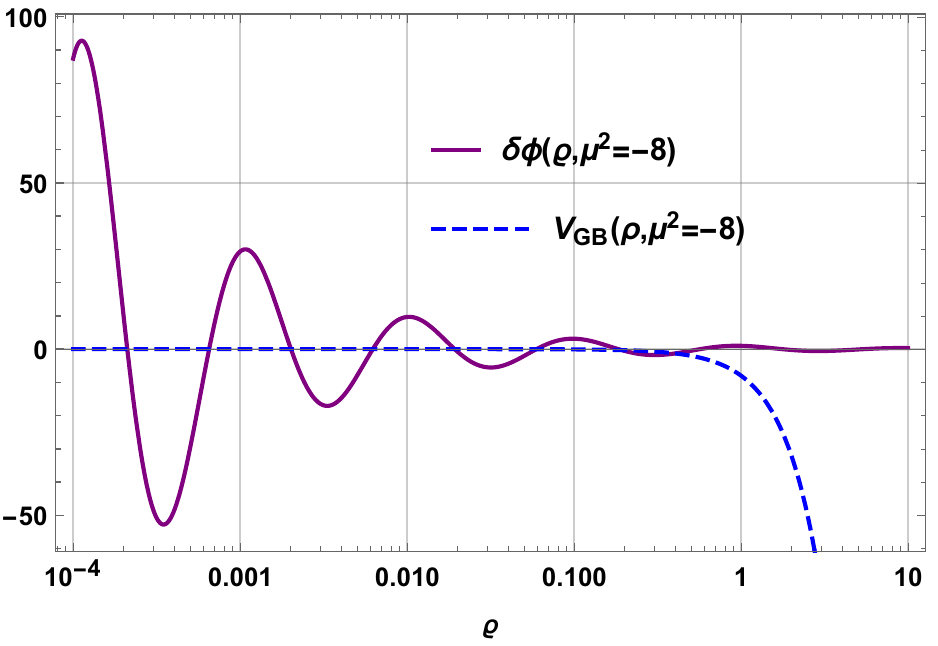}
  (b) \includegraphics[width=0.27\textwidth]{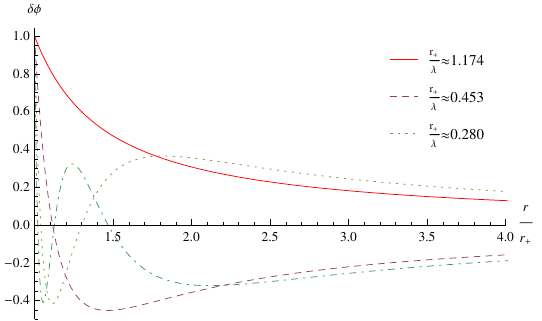}
 (c)\includegraphics[width=0.27\textwidth]{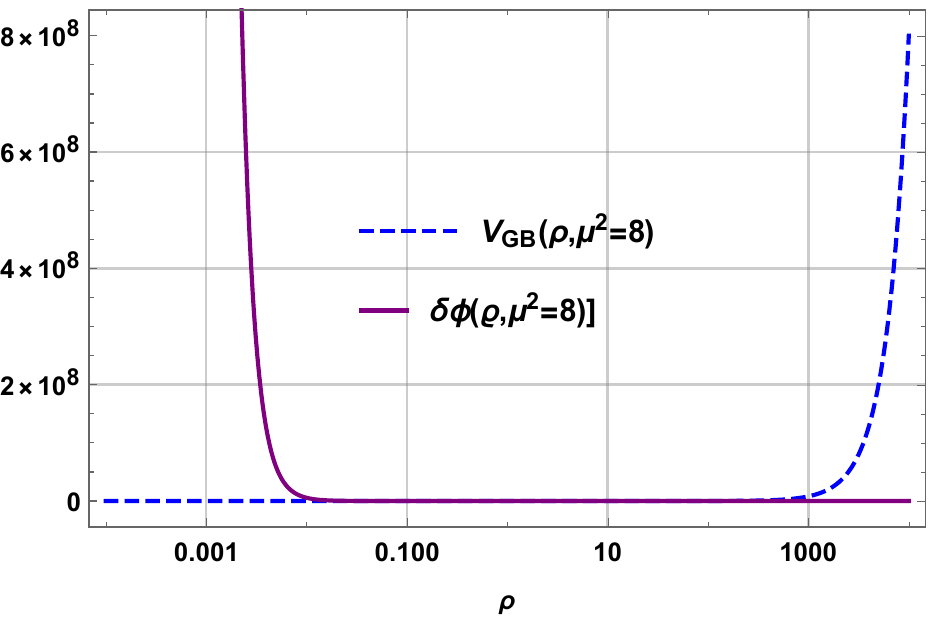}}
\caption{ Three different scalar clouds. (a) Tachyonic  scalar cloud of $\delta \phi(\rho,\mu^2=-8)$ and its negative  potential $V_{\rm  GB}=-8\rho^2$ with negative $\lambda$  for GB$^{\rm BR}$ scalarization.
 This has  many nodes but it has a large value as 100 at $\rho=10^{-4}$.
  (b) Regular scalar clouds  $\delta \phi_n(r,r_+=2)$ with $n=0,1,2,3$ for GB$^+$ scalarization~\cite{Myung:2018iyq}. Here, $n$ represents number of nodes (number of zero-crossings at $r$-axis).
  (c) Scalar cloud of $\delta \phi(\rho,\mu^2=8)$ which blows up  as $\rho\to 0$  and its positive  potential $V_{\rm  GB}=8\rho^2$ with positive $\lambda$.  }
\end{figure}

Consequently, we find two scalar clouds for two branches of negative and positive $\lambda$, showing a feature of GB$^{\rm BR}$ scalarization for cqOSe-black holes.
We notify that the positive branch arose from the BR (AdS$_2\times S^2$) geometry.

\section{Entropy function approach}
Now,  we  perform entropy function approach~\cite{Sen:2005wa}  because the entropy is only the physical quantity to describe scalarized cqOSe black holes~\cite{Marrani:2017uli,Marrani:2022hva}.
The entropy function for the magnetically charged case is defined as~\cite{Myung:2007an}
\begin{equation} \label{E1}
\mathcal{E}=-2\pi \int_{S^2} d\theta d\varphi \Big\{ \sqrt{-g} \mathcal{L}_{\rm EGBS-N} \Big\},
\end{equation}
where its metric  and horizon quantities are given by
\begin{equation}
ds^2_{\rm br}=g_{\mu\nu}dx^\mu dx^\nu=v_1\Big(-\rho^2d\tau^2+\frac{d\rho^2}{\rho^2}\Big)+ v_2(d\theta^2+\sin^2 \theta d\varphi^2), \quad \phi=u,\quad  F_{\theta\varphi}=p \sin \theta.
\end{equation}
Here, $v_1$ and $v_2$ are two parameters  to be determined.
In this case, the entropy function leads to
\begin{equation}\label{E2}
\mathcal{E}=\pi \Big[v_2-v_1+\frac{3v_1 \alpha p^2}{v_2^2}+8\Big(\lambda u^2-\zeta u^4\Big)\Big].
\end{equation}
By extremizing this entropy function with respect to $v_1$ and $v_2$, we find two relations
\begin{equation}
v_2=\sqrt{3\alpha}p=r_e^2,\quad v_1=\frac{\sqrt{3\alpha} p}{2}=\frac{v_2}{2}.
\end{equation}
After eliminating $v_1$ from  the entropy function in Eq.(\ref{E2}), one has the reduced form
\begin{equation}\label{E3}
\mathcal{E}_r=\pi \Big[v_2+8\Big(\lambda u^2-\zeta u^4\Big)\Big],
\end{equation}
which is the same form as that derived by the Wald entropy for $\phi=u$
\begin{equation}
S_{\rm Wald}=-2\pi \oint\frac{\partial \mathcal{L}_{\rm EGBS-N}}{\partial R_{\mu\nu\rho\sigma}} \epsilon_{\mu\nu}\epsilon_{\rho\sigma}r^2d\Omega^2_2|_{r=r_e}.
\end{equation}
Extremizing $\mathcal{E}_r$ with respect to scalar field $u$ leads to
\begin{equation} \label{s-u}
u=\sqrt{\frac{\lambda}{2\zeta}},
\end{equation}
which implies that $\phi=u$ is a secondary hair because it is determined by known coupling constants $\lambda$ and $\zeta$ but not a new scalar charge $q_s$.
Also, to get a real $u$, one requires the same sign for $\lambda$ and $\zeta$.
Plugging Eq.(\ref{s-u})  into Eq.(\ref{E3}), we find the final entropy function
\begin{equation}\label{E4}
\mathcal{E}_f(\alpha,p,\lambda,\zeta)=\pi \Big(v_2+\frac{2\lambda^2}{\zeta}\Big)\to \pi \Big(\sqrt{3\alpha} p+\frac{2\lambda^2}{\zeta}\Big).
\end{equation}
\begin{figure}
\centering
\includegraphics[angle =0,scale=0.5]{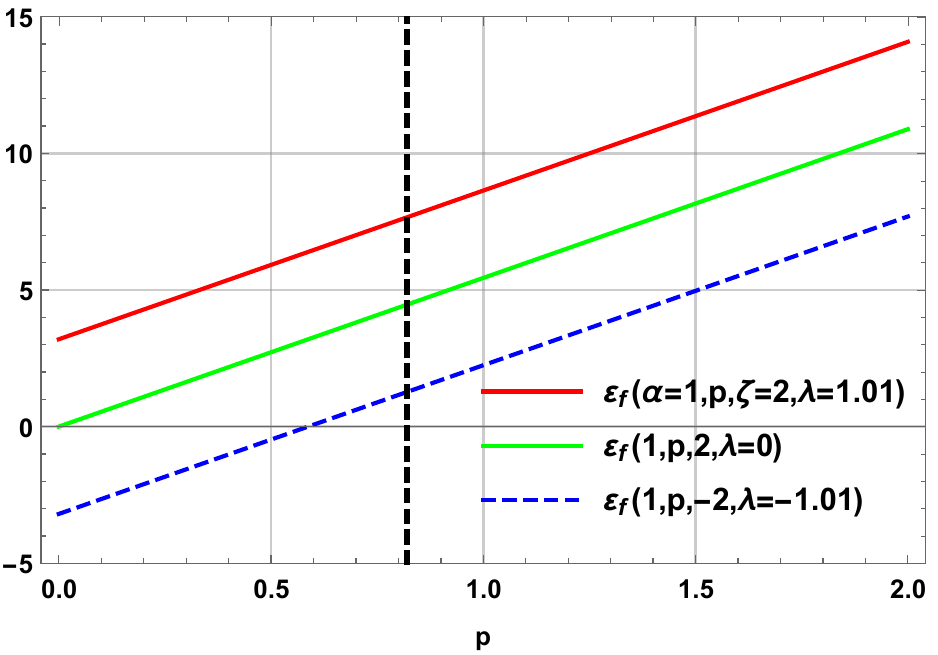}
\caption{ Entropy function $\mathcal{E}_f(\alpha=1,p,\zeta=\pm 2,\lambda)$ with $\lambda=1.01,0,-1.01$  as a function of magnetic charge $p$ for scqOSe- and cqOSe-black holes.
The dashed line is located at $p=0.82$. For $p=0.82$, one has $\mathcal{E}_f(1,0.82,2,1.01)=7.658$, 4.462 for $\lambda=0$, and $\mathcal{E}_f(1,0.82,-2,-1.01)$=1.266.   }
\end{figure}
As is shown in Fig. 5, we have a sequence of entropy: $\mathcal{E}_f(\alpha=1,p,\zeta= 2,\lambda=1.01)>\mathcal{E}_f(1,p,2,0)>\mathcal{E}_f(1,p,- 2,-1.01)$.
This  implies that a phase transition from cqOSe to scalarized cqOSe with $\lambda=1.01$ may occur.
For $p=0.82$, one has $\mathcal{E}_f=7.658(\lambda=1.01)$, 4.462 for $\lambda=0$, and $\mathcal{E}_f(1,0.82,-2,-1.01)$=1.266. where the first/last correspond to the entropy for two scalar clouds (two branches) of $\mu^2=\pm 8$ in Fig. 4(a, c). For $\lambda=-1.01$, one finds that the positive entropy of  $\mathcal{E}_f(1,p,-2,-1.01)>0$ is obtained for $p>0.5873$.

\section{Discussions}
 We have   investigated GB$^-$ scalarization of cqOSe-black hole in the EGBS theory with the NED term $\mathcal{L}_{\rm NED}$.
    This extremal black hole is described by   quantum  $\alpha$  and magnetic charge $P$.  Importantly, this is equivalent to the qOS-extremal black hole whose action ($\mathcal{L}_{\rm qOS}$) is still unknown  when  imposing the relation between charge and mass [$(3\alpha P^2)^{1/4}\to 3M/2$].

    Studying on the onset of GB$^-$ scalarization based on the linearized theory with coupling parameter $\lambda$, we expected to find the single branch of scalarized cqOS extremal (scqOSe)-black holes.
     However, we could not find its scalar cloud which is a seed to generate the single branch. To obtain  scalar clouds, we consider its  near-horizon geometry of the BR spacetime (AdS$_2\times S^2$).
     Two scalar clouds were found to represent two branches: One branch with $\lambda=-1.01$ is represented   the tachyonic cloud with many nodes (but it takes a large value of 100 at the horizon $\rho=10^{-4}$).  The other branch  with $\lambda1.01$ is represented by  the regular cloud without node (but it  blows up at the horizon). The latter arose from the BR (AdS$_2\times S^2$) geometry.

    We have applied  Sen's entropy function approach to this theory with quartic coupling function $f(\phi)=2(\lambda \phi^2-\zeta \phi^4)$. If $\zeta=0$, one cannot obtain scalar hair $u\not=0$.   We obtained the entropy function which shows  the only physical quantity to describe the scqOSe-black holes. The $\lambda=\pm1.01$ branches induce different entropy, leading to that $\lambda=1.01 $ branch is preferred  than  $\lambda=-1.01$ branch. Furthermore, we found that $\mathcal{E}_f(\alpha=1,p,\zeta=2,\lambda=1.01)$ is greater than $\mathcal{E}_f(1,p,2,\lambda=0)$. This  implies that a phase transition from cqOSe to scalarized cqOSe
    may occur.

\section{Acknowledgments}

The author  is supported by the National Research Foundation of Korea (NRF) grant
 funded by the Korea government (MSIT) (RS-2022-NR069013).

\end{document}